\documentstyle[12pt,aps,prl,epsf]{revtex}

\bibstyle{unsrt}
\begin{document}
\tighten
\draft 


\title{Quantum-Mechanical Carnot Engine}

\author{Carl M. Bender$^1$, Dorje C. Brody$^2$, and Bernhard K. 
Meister$^3$} 
\address{${}^1$Department of Physics, Washington University, St. 
Louis MO 63130, USA}
\address{${}^2$Blackett Laboratory, Imperial College, London SW7 
2BZ, UK\\ 
and Department of Applied Mathematics and Theoretical Physics,
University of Cambridge, Silver Street, Cambridge CB3 9EW, UK}
\address{${}^3$Goldman Sachs, Peterborough Court, 133 Fleet 
Street, London EC4A 2BB, UK}

\date{\today}
\maketitle

\begin{abstract}
A cyclic thermodynamic heat engine runs most efficiently if it is 
reversible. Carnot constructed such a reversible heat engine by 
combining adiabatic and isothermal processes for a system containing 
an ideal gas. Here, we present an example of a cyclic engine based on 
a single quantum-mechanical particle confined to a potential well. 
The efficiency of this engine is shown to equal the Carnot efficiency 
because quantum dynamics is reversible. The quantum heat engine has a 
cycle consisting of adiabatic and isothermal quantum processes that 
are close analogues of the corresponding classical processes.
\end{abstract}

\vskip .5cm

\section {INTRODUCTION}
\label{s1}

A thermodynamic heat engine converts heat energy into mechanical work 
by means of a gas that expands and pushes a piston in a cylinder. A 
heat engine obtains its energy from a high-temperature heat reservoir. 
Some of the energy taken from this reservoir is converted to useful 
mechanical work. However, because a heat engine is not perfectly 
efficient some of the energy taken from the heat reservoir is 
{\it not} converted to mechanical energy, but rather is transferred 
to a low-temperature reservoir.

The efficiency of a heat engine is defined as follows \cite{THERMO}: 
If an amount of heat energy $Q_H$ is taken from the high-temperature 
reservoir and an amount $W$ of mechanical work is obtained, then the 
efficiency $\eta$ of the heat engine is defined to be
\begin{eqnarray}
\eta={W\over Q_H}.
\label{e1}
\end{eqnarray}
The amount of heat deposited in the low-temperature reservoir is $Q_C$,
which by conservation of energy is given by
\begin{eqnarray}
Q_C=Q_H-W.
\label{e2}
\end{eqnarray}

It is easy to show that a heat engine running between a given pair 
of high-temperature and low-temperature reservoirs achieves maximum 
efficiency if it is {\it reversible}. Of course, in practice it is 
impossible to construct a heat engine that is perfectly reversible. 
However, Carnot, in the early 19th century, proposed an ideal 
mathematical model of a heat engine that is not only reversible but 
also cyclic \cite{CARNOT}. The Carnot engine consists of a cylinder 
of ideal gas that is alternately placed in thermal contact with the 
high-temperature and low-temperature heat reservoirs whose 
temperatures are $T_H$ and $T_C$, respectively. Carnot showed that 
the efficiency $\eta$ of such a reversible heat engine is
\begin{eqnarray}
\eta=1-{T_C\over T_H}.
\label{e3}
\end{eqnarray}

The Carnot cycle consists of four processes, each of which is 
reversible. First, the gas in the cylinder undergoes an isothermal 
expansion at temperature $T_H$ while it is in contact with the 
high-temperature reservoir. Second, the gas continues to expand 
adiabatically in thermal isolation until its temperature drops to 
$T_C$. Third, the gas is compressed isothermally in contact with the 
low-temperature reservoir. Fourth, the gas is compressed 
adiabatically until its temperature rises to $T_H$.

In this paper we construct an idealised reversible heat engine that 
consists of a single quantum-mechanical particle contained in a 
potential well. Rather than having an ideal gas in a cylinder, we 
allow the walls of the confining potential to play the role of the 
piston by moving in and out. We show that there exist the quantum 
equivalents of isothermal and adiabatic reversible thermodynamic 
processes. However, in place of the temperature variable in 
classical thermodynamics, we use the energy as given by the 
pure-state expectation value of the Hamiltonian. Not surprisingly, 
since this engine is a reversible engine, its efficiency is identical 
to the classical result in Eq.~(\ref{e3}) with $T$ replaced by the 
expectation value of the Hamiltonian.

In our formulation, we do not use the concept of temperature. In a 
classical thermodynamic system, such as an ideal gas in a cylinder, 
the temperature is determined by the average velocity of the large 
number of gas molecules. The system we discuss here is a single 
quantum particle in a potential well. To interpret such a system 
we think of having an infinite number of copies of such particles, 
each in its own potential well. We replace the role of temperature, 
which is an average, by the expectation value of the Hamiltonian, 
that is, the ensemble average of the energies of the quantum 
particle. Thus, the quantum analogue of keeping our quantum 
ensemble in contact with a heat bath as the walls of the potential 
move is to maintain, by some unspecified physical means, the 
constancy of the expectation value of the Hamiltonian. 

This paper is organised as follows: In Sec.~\ref{s2} we describe 
various kinds of quantum processes that may be undergone by a 
particle confined to a potential well and we explain how to perform 
reversible quantum expansions and compressions. Next, in 
Sec.~\ref{s3} we combine these processes to formulate a reversible 
cyclic quantum heat engine. We show that this quantum heat engine is 
the precise analogue of a classical heat engine running on a 
{\it monatomic} ideal gas. Finally, in Sec.~\ref{s4} we make some 
general remarks concerning quantum heat engines and discuss some 
differences between classical and quantum heat engines.

\section {ADIABATIC, FREE, AND ISOTHERMAL QUANTUM PROCESSES}
\label{s2}

Let us consider a particle of mass $m$ confined to a one-dimensional 
infinite square well of width $L$. The time-independent Schr\"odinger 
equation for this system is 
\begin{eqnarray}
-{\hbar^2\over2m}\psi''(x)-E\psi(x)=0,
\label{e4}
\end{eqnarray}
where $\psi(x)$ is required to satisfy the boundary conditions 
$\psi(0)=0$ and $\psi(L)=0$. A general solution to this equation can 
be expressed as a linear combination of eigenfunctions $\phi_n(x)$,
\begin{eqnarray}
\psi(x)=\sum_{n=1}^{\infty}a_n\phi_n(x),
\label{e5}
\end{eqnarray}
where the coefficients $a_n$ satisfy the normalisation condition
\begin{eqnarray}
\sum_{n=1}^{\infty}|a_n|^2=1.
\label{e6}
\end{eqnarray}
The normalised eigenstates $\phi_n(x)$ of this system are
\begin{eqnarray}
\phi_n(x)=\sqrt{2\over L}\sin\left({n\pi\over L}x\right) 
\quad(n=1,2,3,\ldots)
\label{e7}
\end{eqnarray}
and the corresponding eigenvalues $E_n$ are
\begin{eqnarray}
E_n={\pi^2\hbar^2 n^2\over 2mL^2}\quad(n=1,2,3,\ldots).
\label{e8}
\end{eqnarray}

Let us now suppose that one of the infinite walls of the potential 
well, say the wall at $x=L$, can move like the piston in a 
one-dimensional cylinder for a classical thermodynamic system. If 
this wall is allowed to move an infinitesimal amount $dL$, then the 
wave function $\psi(x)$, eigenstates $\phi_n(x)$, and energy levels 
all vary infinitesimally as functions of $L$. As a consequence, the 
expectation value of the Hamiltonian $E(L)=\langle\psi|H|\psi\rangle$ 
also changes infinitesimally. It is natural to define the force on 
the wall of the potential well as the negative derivative of the 
energy. Hence, the force $F$ exerted on the wall of the well is given 
by $F=-{dE(L)\over dL}$.

Based on this force, we can now define several kinds of processes 
which are the quantum analogues of classical thermodynamic processes.

\subsection{Adiabatic Process}
\label{ss2.1}

Classically, an adiabatic process is one in which the system is 
thermally isolated. Thus, for a gas in a cylinder heat cannot flow 
into or out of the gas. In this process the piston moves but the 
system remains in equilibrium at all times. As the piston moves, the 
gas in the cylinder does work. Thus, some of the internal energy of 
the gas is converted into mechanical energy. Let us suppose that the 
gas in the cylinder is a monatomic one-dimensional ideal gas. The 
equation of state of this gas is $PV=NkT$ and the internal energy $U$ 
of the gas is $U={1\over2}NkT$. The mechanical work $dW$ done by an 
infinitesimal expansion $dV$ of the gas is given by $dW=PdV$. By 
definition, in an adiabatic expansion $dU+dW=0$. Solving this 
differential equation gives the standard result that characterises an 
adiabatic process:
\begin{eqnarray}
PV^3=C,
\label{e9}
\end{eqnarray}
where $C$ is a constant.

We assume that the initial quantum state $\psi(x)$ of a particle in a 
square well of width $L$ is a linear combination of eigenstates as in 
Eq.~(\ref{e5}). In an adiabatic process, the size of the potential 
well changes as the wall moves. Since the system remains in 
equilibrium at all times, the absolute values of the expansion 
coefficients $|a_n|$ must remain constant \cite{BF}. That is, we do 
not expect any transitions between states to occur during an 
adiabatic process.  However, it is clear from Eqs.~(\ref{e7}) and 
(\ref{e8}) that as $L$ changes, the eigenstates $\phi_n(x)$ and 
corresponding energy eigenvalues $E_n$ all vary smoothly as functions 
of $L$.

Each energy eigenvalue decreases as the piston moves out (as $L$ 
increases), so the expectation value of the Hamiltonian
\begin{eqnarray}
E(L)=\sum_{n=1}^{\infty}|a_n|^2E_n,
\label{e10}
\end{eqnarray}
where $E_n$ is given in (\ref{e8}), also decreases as a function of 
$L$. The energy that is lost equals the mechanical work done by the 
force $F$, which is given by
\begin{eqnarray}
F(L)=\sum_{n=1}^{\infty}|a_n|^2{\pi^2\hbar^2n^2\over mL^3}.
\label{e11}
\end{eqnarray}

Note that during an adiabatic process, the phase of the wave function 
of the particle changes, even though the particle remaines fixed in a 
given eigenstate. This, however, does not affect the results in this 
paper, because we are only interested in expectation values which are 
independent of phases. 

\subsection{Free Expansion}
\label{ss2.2}

Classically, a free expansion is one in which the piston suddenly 
moves outward. The gas in the cylinder instantly departs from 
equilibrium and expands to fill the new volume. During this process 
the system remains isolated from any heat source, so heat energy 
cannot flow into or out of the gas in the cylinder. Since the piston 
is not pushed outward by the gas, the system does no work. Thus, the 
internal energy in the gas remains constant during this process. The 
temperature of an ideal gas is proportional to the internal energy of 
the gas, and thus  the equation that characterises a free expansion 
process is 
\begin{eqnarray}
T=C,
\label{e12}
\end{eqnarray}
where $C$ is a constant. Also, after the expansion of the gas when 
equilibrium is restored, the final pressure has decreased relative to 
the initial pressure in such a way that $P_{\rm final}V_{\rm final} = 
P_{\rm initial}V_{\rm initial}$.

Quantum mechanically, in a free expansion the initial state $\psi(x)$ 
of the system of volume $L$ becomes a final state for which the 
expectation value of the Hamiltonian is the same as its initial 
expectation value. This is the quantum analogue of the requirement 
that the internal energy remain the same during a free expansion 
because the system does no mechanical work.

At the new value of $L$, say $\alpha L$, where $\alpha>1$, we have a 
new set of eigenfunctions $\chi_n(x)$ of the form
\begin{eqnarray}
\chi_n(x)=\sqrt{2\over\alpha L}\sin\left({n\pi\over\alpha L}x\right)
\quad (n=1,2,3,\ldots).
\label{e13}
\end{eqnarray}
After the expansion, each of the initial eigenstates $\phi_n$ becomes 
a linear combination of the eigenstates $\chi_m(x)$ associated with 
the well of width $\alpha L$:
\begin{eqnarray}
\phi_n=\sum_{m=1}^{\infty}b_{m,n}\chi_m(x),
\label{e14}
\end{eqnarray}
where $b_{m,n}$ is the overlap integral
\begin{eqnarray}
b_{m,n}&=&\int_0^L dx\,\phi_n(x)\chi_m(x)\nonumber\\
&=&{2n\alpha^{3/2}(-1)^n\over\pi(m^2-\alpha^2n^2)}\sin\left(
{m\pi\over\alpha}\right).
\label{e15}
\end{eqnarray}

Using this formula we can verify that the expectation values of the 
Hamiltonian in the initial state and in the final state are indeed 
the same. Suppose that the initial state is in the $n$-th eigenstate. 
Then, we require $$\sum_{m=1}^{\infty}b_{m,n}^2E_m(\alpha L)=E_n(L)$$ 
to hold for an arbitrary integer $n$ and arbitrary $\alpha>1$. Using 
Eq.~(\ref{e15}) for $b_{m,n}$, this can be written as 
\begin{eqnarray}
\sum_{m=1}^{\infty}{4\alpha m^2\over\pi^2(m^2-\alpha^2n^2)^2}\sin^2
\left({m\pi\over\alpha}\right)=1.
\label{e16}
\end{eqnarray}
To verify this identity, we note that it can be rewritten more simply 
as $$\left.\left[{\alpha\over\pi^2}\left(n{\partial\over\partial n} + 
2\right) \sum_{m=1}^\infty{1\over m^2-\alpha^2n^2}\left(1-\cos{2\pi 
m\over\alpha}\right)\right]\right|_{n\,=\,{\rm positive~integer}}=1.$$ 
Then, to establish this equation we use the general formula 
$$\sum_{m=1}^\infty{\cos(mx)\over m^2-u^2}={1\over2u^2}-{\pi\cos\{
[(2k+1)\pi-x]u\}\over2u\sin(\pi u)},$$ 
where $k$ is the largest integer such that $2\pi k\leq x$.\cite{GR} 
Next, we put $x=2\pi/\alpha$ and $u=\alpha n$ and take $k=0$ because 
we have assumed that $\alpha>1$. It is now easy to show that 
Eq.~(\ref{e16}) is satisfied.

Note that in the textbook by Schiff \cite{SCHIFF} it is stated that 
when a quantum system undergoes a sudden expansion, the associated 
energy generally changes. The point that is not stated therein, 
however, is that the {\it expectation value} of the Hamiltonian is 
conserved in such a process, as we have illustrated above.

\subsection{Isothermal Process}
\label{ss2.3}

Classically, an isothermal process is one in which as the piston 
moves, the system remains in equilibrium at all times. During this 
process the system is in contact with a heat source so that the 
temperature $T$ of the gas in the cylinder remains fixed. As the 
piston moves, the system does work. However, since the temperature of 
the gas remains constant, the internal energy of the gas also remains 
constant. Thus, if the equation of state of the gas is $PV=NkT$, then 
the equation that characterises an isothermal process is
\begin{eqnarray}
PV=C,
\label{e17}
\end{eqnarray}
where $C$ is a constant.

Quantum mechanically, we characterise an isothermal expansion as 
follows. We assume that the initial state $\psi(x)$ of the system of 
volume $L$ is a linear combination of eigenstates as in 
Eq.~(\ref{e5}). In an isothermal process, the size of the potential 
well changes as the wall moves. However, the expectation value of the 
Hamiltonian remains constant as the size of the well changes. This 
expectation value is an ensemble average taken over the multiple 
copies of the system. Technically speaking, we regard this ensemble 
as a mixed state; we are not concerned with the phases of the states 
making up the ensemble. (In principle, constancy of the expectation 
value of the Hamiltonian may be achieved by pumping energy into the 
system, possibly by using lasers.) Thus, the expansion coefficients 
$a_n$ must change in such a way as to keep 
\begin{eqnarray}
E(L)=\sum_{n=1}^{\infty}|a_n|^2E_n,
\label{e18}
\end{eqnarray}
fixed as $L$ increases.

The instantaneous force on the piston as a function of $L$ is given 
by Eq.~(\ref{e11}). However, unlike the adiabatic case, the 
coefficients $|a_n|^2$ in (\ref{e11}) are now no longer constants, 
but rather vary as functions of $L$ subject to the constraint that
\begin{eqnarray}
\sum_{n=1}^\infty |a_n|^2=1.
\label{e19}
\end{eqnarray}

\section{QUANTUM CARNOT CYCLE}
\label{s3}

Using the quantum adiabatic process and the quantum isothermal 
process described in Sec.~\ref{s2} we can now construct a cyclic heat 
engine. Note that we cannot use a quantum free expansion because this 
process is not reversible. We consider first a particularly 
simple case in which only {\it two} of the eigenstates of the 
potential well contribute to the wave function in the well. Then, we 
consider the general case in which any number of states participate.

\subsection{Two-State Quantum Heat Engine}
\label{ss3.1}

We consider the following cyclic process. We start with a 
ground-state wave function in a well of width $L_1$. At 
this point the force on the wall is 
$$F={\pi^2\hbar^2\over mL_1^3}$$
and the expectation value of the Hamiltonian is 
\begin{eqnarray}
E_H={\pi^2\hbar^2\over2mL_1^2},
\label{e20}
\end{eqnarray}
where the subscript $H$ is in analogy with the $T_H$ in 
Eq.~(\ref{e3}). 

{Step I}: We allow the piston to expand isothermally. As we do so, we 
excite the second energy level of the system keeping the expectation 
value of the Hamiltonian constant. Thus, during this isothermal 
expansion, the state of the system is a linear combination of the 
lowest two energy eigenstates: 
\begin{eqnarray}
\psi(x)=a_1(L)\sqrt{2\over L}\sin\left({\pi\over L}x\right)
+a_2(L)\sqrt{2\over L}\sin\left({2\pi\over L}x\right),
\label{e21}
\end{eqnarray}
where, from Eq.~(\ref{e19}), we have $|a_1|^2+|a_2|^2=1$. The 
expectation value of the Hamiltonian in this state as a function of 
$L$ is 
$${\pi^2\hbar^2\over2mL^2}(4-3|a_1|^2),$$
where we have used the condition $|a_1|^2+|a_2|^2=1$ to eliminate 
$|a_2|^2$. Setting the expectation value of the Hamiltonian equal 
to $E_H$ gives 
$$L^2=L_1^2(4-3|a_1|^2).$$
Thus, the maximum possible value of $L$ in this isothermal expansion 
is $L_2=2L_1$, and this is achieved when $a_1=0$. Note that at 
$L=L_2$ the system is purely in the second energy eigenstate. Along 
this isothermal expansion the force as a function of $L$ is
\begin{eqnarray}
F_1(L)=|a_1|^2{\pi^2\hbar^2\over mL^3}+(1-|a_1|^2)
{4\pi^2\hbar^2\over mL^3}={\pi^2\hbar^2\over mL_1^2L}.
\label{e22}
\end{eqnarray}
Observe that the product $LF_1(L)$ is a constant. This is the exact 
quantum analogue of the classical equation of state in 
Eq.~(\ref{e17}). In general, for a one-dimensional quantum system 
undergoing an isothermal process, we have
\begin{eqnarray}
LF(L)=C,
\label{e23}
\end{eqnarray}
where $C$ is a constant.

{Step II}: Next, we allow the system to expand adiabatically from 
$L=L_2$ until $L=L_3$. During this expansion the system remains in 
the second state and the expectation value of the Hamiltonian is
\begin{eqnarray}
E_C={2\pi^2\hbar^2\over mL^2}.
\label{e24}
\end{eqnarray}
The force as a function of $L$ can be taken from Eq.~(\ref{e11}):
\begin{eqnarray}
F_2(L)={4\pi^2\hbar^2\over mL^3}.
\label{e25}
\end{eqnarray}
The product $L^3F_2(L)$ is a constant. This is the quantum analogue 
of the classical equation of state in Eq.~(\ref{e9}). In general, 
for a one-dimensional quantum system undergoing an adiabatic process, 
we have
\begin{eqnarray}
L^3F(L)=C,
\label{e26}
\end{eqnarray}
where $C$ is a constant.

{Step III}: Third, we compress the system isothermally from $L=L_3$ until
$L=L_4$. During this compression we extract energy so that the expectation
value of the Hamiltonian remains constant. At the beginning of the compression
the system is in its second excited state. We choose to compress the system
until it is back in the ground state. At this point $L_4={1\over2}L_3$.
During this compression the expectation value of the Hamiltonian is kept
constant at the value
$${2\pi^2\hbar^2\over mL_3^2}.$$
The applied force as a function of $L$ is
\begin{eqnarray}
F_3(L)={4\pi^2\hbar^2\over mL_3^2L}.
\label{e27}
\end{eqnarray}

{Step IV}: Finally, we continue to compress the system adiabatically from
$L=L_4$ until we return to the starting point $L=L_1$. During this compression
the system remains in the ground state and the expectation value of the
Hamiltonian is
$${\pi^2\hbar^2\over2mL^2}.$$
The force applied to the wall of the potential well is given by
\begin{eqnarray}
F_4(L)={\pi^2\hbar^2\over mL^3}.
\label{e28}
\end{eqnarray}

The four-step cyclic quantum process that we have just described is 
illustrated in fig.~1. The Carnot cycle is drawn in the 
$(F,L)$-plane, which is the one-dimensional version of the 
$(P,V)$-plane. The area of the closed loop represents the mechanical 
work $W$ done in a single cycle of the quantum heat
engine. To calculate $W$ we evaluate the following integrals:
\begin{eqnarray}
W&=&\int_{L_1}^{2L_1}dL\,F_1(L)+\int_{2L_1}^{L_3}dL\,F_2(L)
+\int_{L_3}^{L_3/2}dL\,F_3(L)+\int_{L_3/2}^{L_1}dL\,F_4(L)\nonumber\\
&=&{\pi^2\hbar^2\over m}\left({1\over L_1^2}-{4\over L_3^2}\right)\ln 2.
\label{e29}
\end{eqnarray}

To determine the efficiency of this two-state heat engine we need to calculate
the energy absorbed by the potential well during the isothermal expansion. This
quantity of energy $Q_H$ is given by
\begin{eqnarray}
Q_H&=&\int_{L_1}^{2L_1}dL\,F_1(L)\nonumber\\
&=&{\pi^2\hbar^2\over L_1^2m}\ln 2.
\label{e30}
\end{eqnarray}
Therefore, the efficiency $\eta$ of our two-state quantum heat engine, as given
in the general formula (\ref{e1}), is
\begin{eqnarray}
\eta=1-4\left({L_1\over L_3}\right)^2.
\label{e31}
\end{eqnarray}
Using Eqs.~(\ref{e20}) and (\ref{e24}), we can rewrite this formula as
\begin{eqnarray}
\eta=1-\left({E_C\over E_H}\right).
\label{e32}
\end{eqnarray}
The energy in this formula plays the role of temperature in Eq.~(\ref{e3}), and
thus (\ref{e32}) is the analogue of the classical thermodynamical result of
Carnot.


\subsection{Multiple-State Quantum Heat Engine}
\label{ss3.2}

It is straightforward to generalise the two-state quantum Carnot engine to $n$
states. The only change that occurs is that the maximum length of the 
isothermal
expansion increases. Specifically, we find that for $n$ states, if the initial
width of the well is $L_{1}$, then the maximum value of $L_2$ is bounded above
by $nL_1$.

We note that the entropy $S$ of the quantum system can be defined as 
\begin{eqnarray} 
S=-\sum_{i=1}^{\infty}|a_{i}|^2\ln|a_{i}|^2, 
\label{e33} 
\end{eqnarray} 
when we drop the constraint to limit the allowable number of states. In this
case, the entropy increases during an isothermal expansion, while $S$ decreases
for an isothermal compression. On the other hand, for an adiabatic process, the
entropy remains fixed because the absolute values $|a_{i}|$ of the expansion
coefficients do not change. 

\section{DISCUSSION}
\label{s4}

We have shown that the efficiency of a quantum Carnot cycle is the 
same as that of a classical Carnot cycle, with the identification of 
the expectation value of the Hamiltonian as the temperature of the 
system. However, there is a distinct advantage of a quantum Carnot 
cycle over a classical Carnot cycle. The heat processes of a 
classical Carnot cycle must be reversible and therefore the ideal 
gas in the cylinder must be in equilibrium at all times. If the 
piston moves with finite speed then the gas is immediately out of 
equilibrium and the process is no longer reversible. Therefore, it 
takes an infinitely long time for a classical Carnot engine to 
complete one cycle. 

By contrast, for a quantum Carnot engine, any given cycle lasting 
a finite time $t>0$ has a calculable probability of being reversible. 
For a quantum cycle that lasts an infinite amount of time, during 
the adiabatic part of the cycle, a particle in a given energy 
eigenstate remains in that eigenstate. This is true because of the 
quantum adiabatic theorem \cite{BF,SCHIFF}. If the adiabatic portion 
of the cycle lasts a long but finite amount of time, then there is a 
small but nonzero probability amplitude that the particle may end up 
in a different energy eigenstate. Thus, there is a large probability 
that Carnot cycle is reversible. 

To understand the calculation that must be performed, consider step 
II (the adiabatic quantum expansion) of a two-state Carnot engine as 
described in Sec.~\ref{s3}. For this process, there is a finite 
probability that there will be no transitions between states during 
this process. (The specific calculation of this probability is 
described in Ref.~\cite{SCHIFF} and is not of particular interest 
here.) This probability becomes smaller as the speed of the engine 
increases, but for a finite-speed quantum Carnot engine, we can make 
the speed sufficiently slow that this probability becomes negligibly 
different from unity. Thus, quantum mechanics makes a maximally 
efficient heat engine an actual physical possibility. Furthermore, 
because a quantum Carnot engine is reversible, if it follows the 
cycle in Fig.~1 in reverse, it functions as a quantum 
refrigerator. While a classical refrigerator is inherently 
inefficient, an efficient quantum refrigerator might actually be 
feasible.

Finally, we must address the question of whether a quantum Carnot 
cycle is truly cyclic. In the calculations we have done concerning 
energy, work, force, and efficiency, only the absolute values $|a_n|$ 
of the expansion coefficients in Eq.~(\ref{e5}) appear. In the 
quantum Carnot cycle pictured in Fig.~1 the values of $|a_n|$ 
are indeed cyclic. However, over the course of a cycle the relative 
phases of these coefficients can change; each coefficient may accrue 
a geometric phase contribution. We do not know whether the relative 
phases have physically observable consequences with regard to the 
functioning of a quantum heat engine.

\section* {ACKNOWLEDGEMENTS}
\label{s6}

CMB thanks the Theoretical Physics Group at Imperial College, London, 
for its hospitality. DCB thanks the Royal Society for financial 
support. DCB and BKM thank the Physics Department at Washington 
University for its hospitality. This work was supported in part by 
the U.S.~Department of Energy.

\end{document}